\documentclass{JINST}
\usepackage{graphicx}

\title{Characterisation of a Composite LEPS}

\author{  Moumita Roy Basu$^a$, Sudatta Ray$^b$\thanks{Presently at
Amity   University,   Noida    -    201303,    INDIA}~,    Abhijit
Bisoi$^b$\thanks{Presently   at  Indian  Institute  of  Engineering
Science  and  Technology,  Howrah  -  711103,  INDIA},  M.   Saha
Sarkar$^b$\thanks{Corresponding author.}\\
\llap{$^a$}University of Calcutta,\\ Kolkata - 700073, INDIA\\
\llap{$^b$}Saha  Institute  of  Nuclear  Physics,  \\  Bidhannagar,
Kolkata - 700064, INDIA\\
  E-mail: \email{maitrayee.sahasarkar@saha.ac.in}}

\abstract{A  low  energy  photon  spectrometer (LEPS), which is a
composite planar HPGe, has been characterised experimentally.  It
has  been  shown  that  beyond  200  keV, effect of image charges
deteriorates the efficiency of the detector in its addback  mode.
Data  have  been  corrected  on event-by-event basis resulting in
improvement of the performance.}


\begin{document}

\section{Introduction}

Apart  from  a few medium sized array \cite{afro}, gamma detector
arrays usually have very low  detection  efficiency  for  photons
with  energies  less  than  50 keV due to the absence of suitable
detectors which are efficient at  these  low  energies.  However,
this  energy  range  has  special importance for detection of (i)
X-rays for identification  of  fission-  fragments,  (ii)  highly
converted  low  energy  gammas emitted from isomers in medium and
heavy nuclei,  (iii)  gamma  transitions  connecting  the  lowest
states  in heavy nuclei, etc. The low energy photon spectrometers
(LEPS) having reasonable resolution at  this  energy  range  have
very  low  efficiency.  So  it  is  desirable  to use a composite
detector which includes more than one  planar  LEPS  having  good
resolution  to  get  higher  efficiency.  Electrically  segmented
planar LEPS has been used in some of the arrays as a solution  to
this  problem.  In  this  paper,  a segmented planar LEPS will be
characterized.

\begin{figure}[htb]
\centering
  \begin{tabular}{@{}cc@{}}
\includegraphics[width=.43\textwidth]{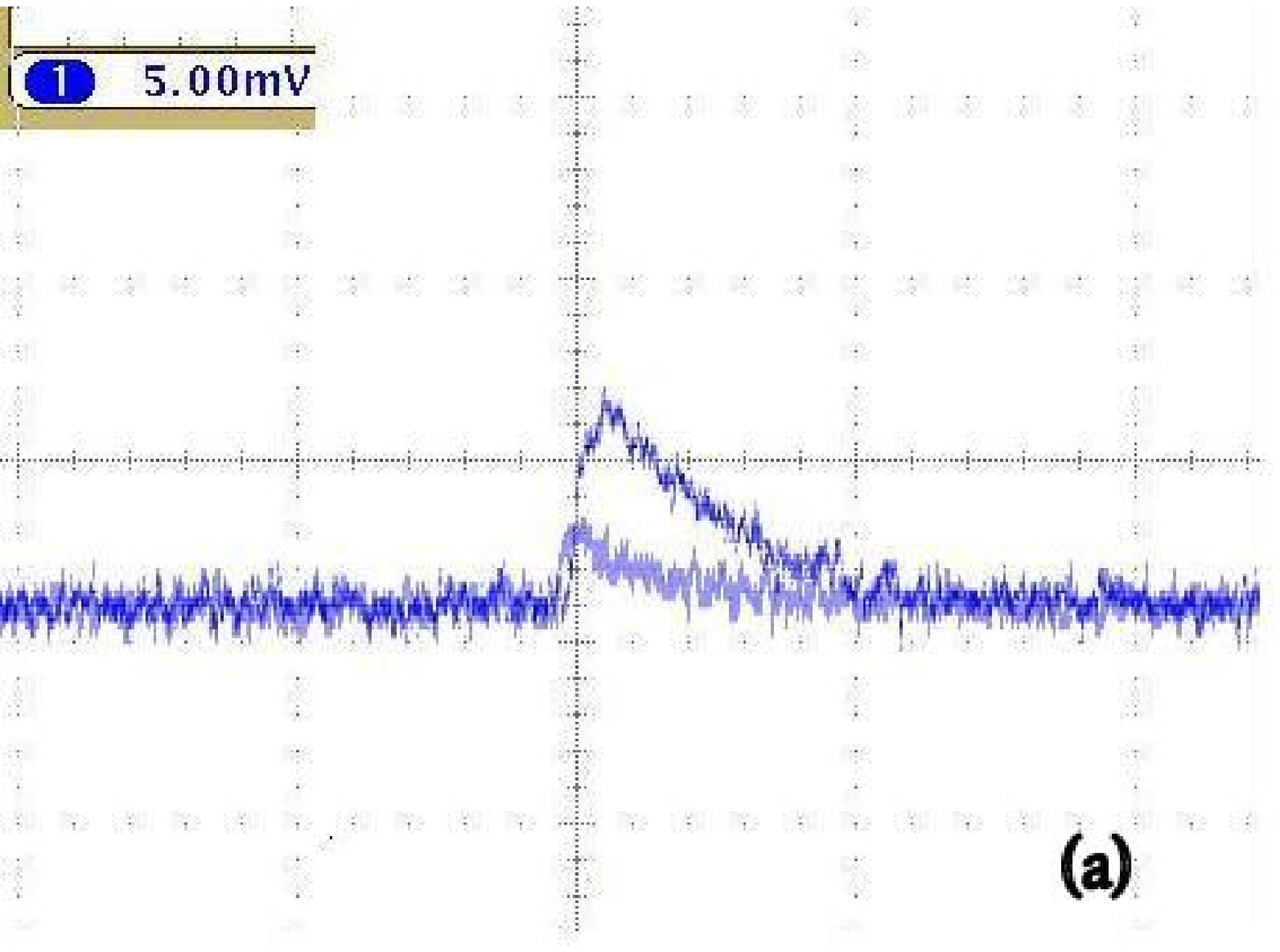} &
    \includegraphics[width=.43\textwidth]{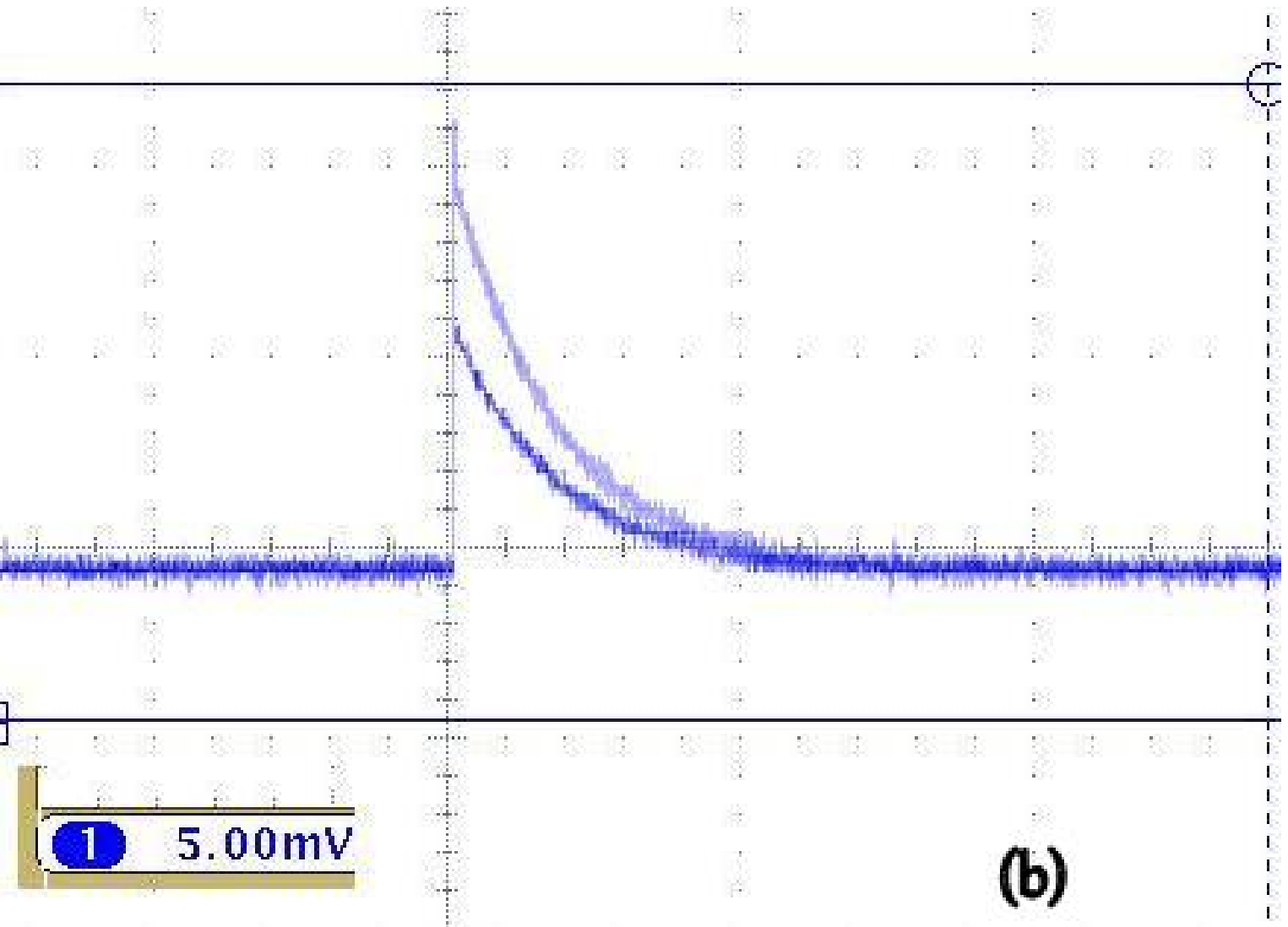}
\end{tabular}
 \caption{
\label{nobias}
Preamplifier outputs of one  of  the  segments  of  the  Composite
detector, (a) without and (b) with bias.}
\end{figure}

\section{Experimental Details }

\subsection{The detector}

In  the  present  work; a composite planar HPGe low energy photon
detector (designated as "Composite" along  the  text)  \cite{dae}
with   four   segments   mounted  on  a  common  cryostat  (Table
\ref{spec}) similarly toa Clover  detector  has  been  used.  The
detector   has   been   fabricated   by   DSG   Detector  Systems
Company\footnote{DSG Detector Systems GmbH, Robert Bosch  Strasse
38,  55129  Mainz, Telephone: 06131 50 75 30, Fax: 06131 5075 40,
email: info [at] detectorsystems.de. The company has pulled  back
from  operations.},  Germany. The crystal in this planar detector
is of p-type. The volume of each crystal (segment) is  such  that
each  of  them has a resolution of around 500 eV for 122 keV. The
efficiency of the total system is around 4 times that of a single
detector. The active area is 4 $\times$ 80 mm$^2$. It has a  thin
Be   window  (Table  \ref{spec}),  to  maximize  the  low  energy
efficiency of the detector. The necessary reverse bias  for  this
composite  LEPS  for  operation  is  very  low, only -300 V. Bias
always depends on the material, crystal quality and design.  This
low  bias  for  optimum  performance  indicates  that the crystal
quality is good. The Composite detector use a single bias supply.
The detector assembly is designed such that the  supply  provides
reverse voltage to four segments.

The noise level is quite low even without bias. Fig. \ref{nobias}
shows  the  preamplifier  output from one of the crystals without
bias as well as the same preamplifier output with -300 V detector
bias.

In  this  work  the  performance  of  this detector with a normal
planar HPGe detector (ORTEC GLP 10180/07 SH  GLP:  designated  as
"Old  LEPS")  will  be  compared.  Specifications  of  these  two
detectors are shown in Table \ref{spec}.

\begin{figure}[ht]
\begin{center}
\vspace{3.5  cm}
\includegraphics[width=.8\linewidth,height=.7\linewidth]
{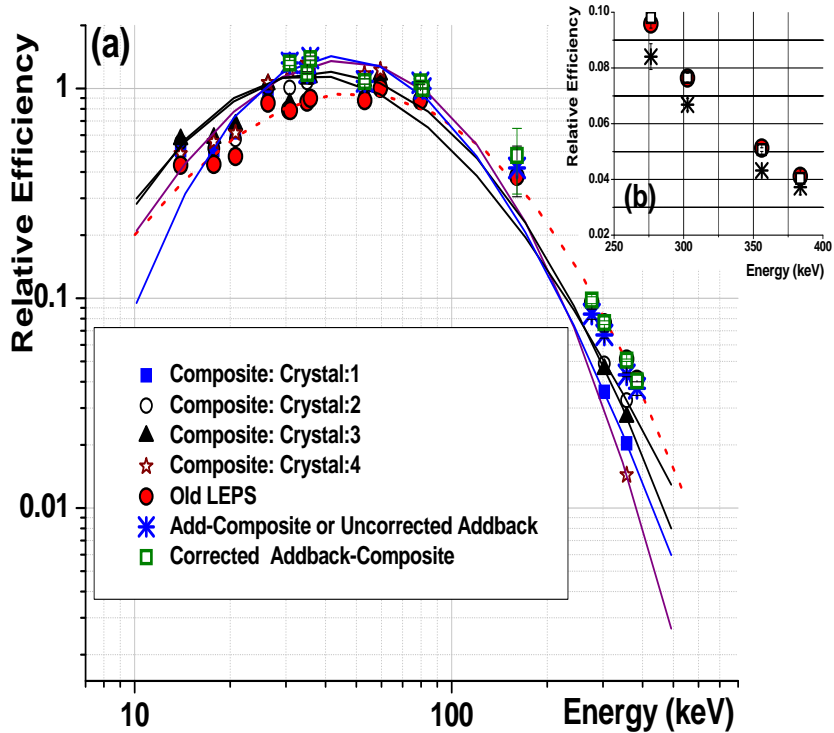}
\vspace{-4.cm}
\caption{  \label{eff} (a) Comparison of relative efficiencies of
individual crystals of Composite LEPS in  singles,  add,  addback
and  corrected  addback  modes  with  that  of  the Old LEPS. (b)
(Inset) The differences between relative efficiencies of Old LEPS
and Composite LEPS, in add and corrected addback modes have  been
displayed in an expanded scale for clarity. The efficiencies have
been normalised to 1 at 81 keV. }

\end{center} \end{figure}

\subsection{Modes of detection in a composite  detector}

The  total  full energy peak efficiency of any composite detector
with more than one crystal (segment) includes the effect  of  two
complimentary processes \cite{mss1}. They are

\begin{itemize} \item{}single fold events, when full gamma energy
is  deposited  in  any  one  of  the individual crystals. This is
related to the direct full energy peak detection efficiencies  of
individual crystals,

\item{}the   full   gamma   energy  resulting  from  the  partial
absorption in two or more crystals through Compton effect (and/or
pair production process followed by escape of one/both of the 511
keV gamma rays for high energy gammas). This is  related  to  the
coincidence detection efficiency. \end{itemize}

Direct  detection  efficiency of a composite detector is obtained
by treating the signals from individual crystals  separately.  It
is  obtained  from  the  uncorrelated sum of spectra from each of
these crystals. On the other hand, in coincidence detection mode,
two or more crystals are in temporal coincidence, mainly  due  to
Compton scattering from one crystal to its neighbouring ones. The
energy  signals  of individual crystals fired simultaneously, are
recorded event  by  event  in  the  list  mode.  The  coincidence
detection  efficiency  is  determined  from  the spectra (addback
spectra) generated from time correlated sum  (addback)  of  these
signals.  So  the  total full energy peak detection efficiency of
this  composite detector is   $\epsilon_{total}   =   \epsilon_{direct}
+\epsilon_{addbk}$. Therefore, at relatively higher energy, where
Compton  processes  become important, the addback contribution is
responsible for the enhancement of the  total  full  energy  peak
detection efficiency compared to the direct mode. As a measure of
this enhancement, the addback factor F is defined as the ratio of
the  total  full  energy  peak efficiency to the direct detection
efficiency; {\it i.e.}, F=$\epsilon_{total}/\epsilon_{direct}$.

\begin{table}[ht]
\caption{Specifications of the two detectors used in the present work}
\label{spec}
\begin{center}
\begin{tabular}{cccccccccc}
\hline
\multispan{3}\hfil  Detector \hfil&& \multispan{3}\hfil Dimensions \hfil&&
\multispan{2} \hfil Be-Window\hfil
\\\\
\multispan{3}\hrulefill && \multispan{3}\hrulefill &&
\multispan{2} \hrulefill
\\
Make&Type&High&&Dia.& Height& Area&& Dist. \footnote{}
& Thick\\
&&Voltage(V)&&(mm)& (mm)& (mm$^2$)&& (mm)
& (mm)\\ \hline
ORTEC&GLP&-1000&&10&7&80&&5&0.127\\
     &Planar: n-type&\\
     &single& \\
\strut \\
DSG&PGP 4seg80-7&-300&&24&7&4$\times$80&&5&0.127\\
&Planar: p-type&\\
     &4 segments& \\
\hline
\end{tabular}\end{center}
{$^2$}{Distance between the crystal and the endcap}
\end{table}

\begin{figure}[htb]
\vspace{4. cm}
\includegraphics[width=\linewidth,height=.6\linewidth] {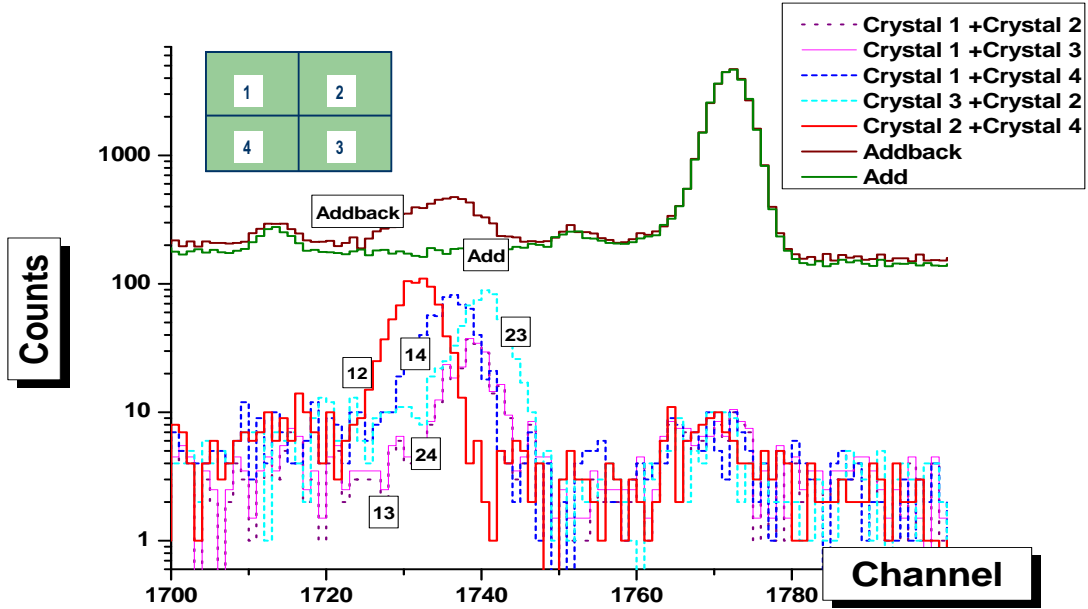}
\vspace{-4.8cm}
\caption{
\label{twofold}
The  comparison  of  add,  addback and different two-fold spectra
(12,14,23,24 and 13)  for  the  Composite  LEPS.  The  two  -fold
spectra  12  (say,)  is  generated  from  simultaneous  events in
segments 1 and 2. The segments numbers are indicated in the inset
figure.} \end{figure}

\subsection{The Experiment and Data Analysis}

We  have  used  $^{152}Eu$, $^{133}Ba$ and $^{241}Am$ radioactive
sources to determine the relative efficiency  of  the  detectors.
The efficiency of the composite detector has been determined from
the  sum  spectrum  generated  by addition of independent singles
spectra of four segments. We have designated this sum spectrum as
$Add$ spectrum and this mode as $Add$ mode. In Clover  detectors,
at  around 1 MeV, the addback improves the absolute efficiency by
about 1.5 times \cite{mss1}. It means that the efficiency in  the
$Addback$  mode  is  1.5  times more than that in the $Add$ mode.
However, this ratio is energy dependent. At low  energies,  where
Compton   effect   is   not  dominant,  addback  process  is  not
advantageous. To take advantage of the addback mode (as discussed
earlier in Section 2.2) at  energies  above  200  keV  where  the
efficiency  of  a  small  area  planar  detector  falls  sharply,
correlated total  spectrum  of  the  four  crystals  (similar  to
addback  spectrum  of a Clover) has been generated. In this case,
the $Addback$ spectrum is generated  by  event  wise  summing  of
correlated  data  obtained  in  four  segments  of  the composite
detector. We have designated this mode as $Addback$ (uncorrected)
in the figures and text. We have  collected  the  data  from  the
composite  detector  in the LIST Mode. In this work, the relative
efficiency of the composite  detector  in  $Add$  and  correlated
$Addback$ modes have been compared with the relative efficiencies
of individual crystals and that of Old LEPS)(Fig.\ref{eff}a).

\section{Results and Discussion}

\subsection{Relative efficiency}

It  has  been found that any single crystal of the composite LEPS
(Fig.\ref{eff}a) or the summed spectra of the four crystals  show
almost  similar relative efficiencies at low energies ($\leq$ 100
keV) when compared with the Old LEPS. It is observed that for the
composite detector, the relative efficiency in the  $Addback$  is
exactly  similar to that in the $Add$ mode. So they are indicated
as $Add-Composite$ (uncorrected Addback) in  Fig.  \ref{eff}.  No
improvement  is  noted  beyond 200 keV. It is even worse than the
Old LEPS (Fig. \ref{eff}b). At higher energies the efficiency  of
the   Old   LEPS   improves   and   is   better  than  the  $Add$
($Addback$-uncorrected) modes of the composite detector. This  is
an   unexpected   observation  and  needs  closer  inspection  to
determine its cause.

\begin{figure}[h]
\vspace{4.5   cm}
\includegraphics[width=\linewidth,height=.6\linewidth]
{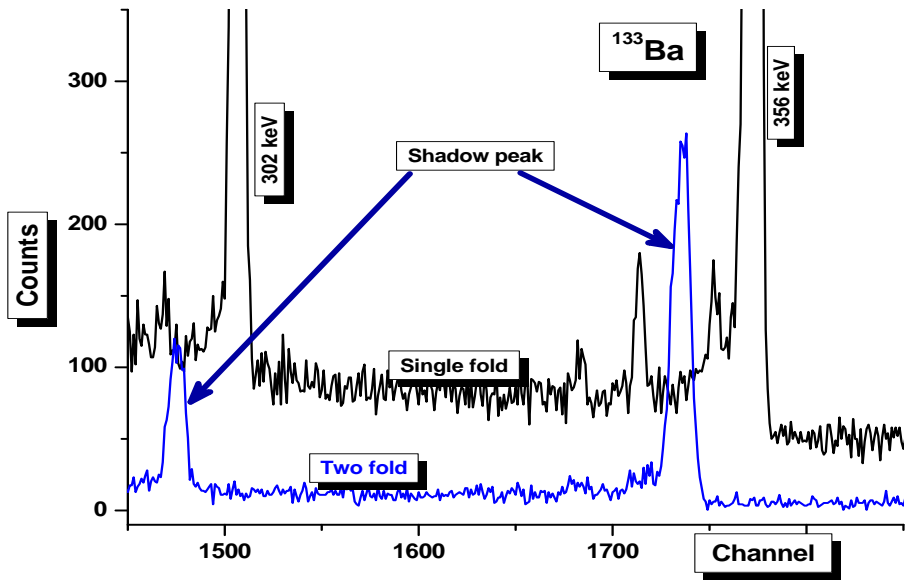}
\vspace{-5.cm}
\caption{  \label{addbk}  The  single-  and  two-fold spectra are
compared for a $^{133}Ba$ source. }
\end{figure}

\subsubsection{Causes}

 The   add,   addback,  single-  fold  and  two  -  fold  spectra
(Figs.\ref{twofold},   \ref{addbk})   have   been   compared   to
understand this feature. It is observed that the addback spectrum
of  the composite LEPS contains a relatively low intensity shadow
peak  corresponding  to  each  higher  energy  gamma  peak  (Fig.
\ref{twofold}).   While  comparing  the  spectra  generated  from
single- fold and two- fold events only, it is observed that these
peaks are solely generated by the two fold events as evident from
Fig.\ref{addbk}. The two -fold  spectrum  (Fig.\ref{addbk})  does
not  contain  the  photopeak,  it  only  includes the shadow peak
corresponding to it.

To  identify  the  origin  of these shadow peaks more accurately,
two- fold spectra generated from neighbouring crystals (12,14 and
23) as well as diagonally opposite  crystals  (13,24)  have  been
compared   in   Fig.\ref{twofold}.   It   is   clearly  seen  the
neighbouring crystals are  contributing  nearly  double  of  that
contributed  by  the diagonally opposite ones in the shadow peak.
The peak-shifts vary as a function  of  second  crystal  position
with  respect  to  that  of  the  first  crystal depending on the
position of the source.

\begin{figure}[h]
\centering
\includegraphics[width=0.8\textwidth]{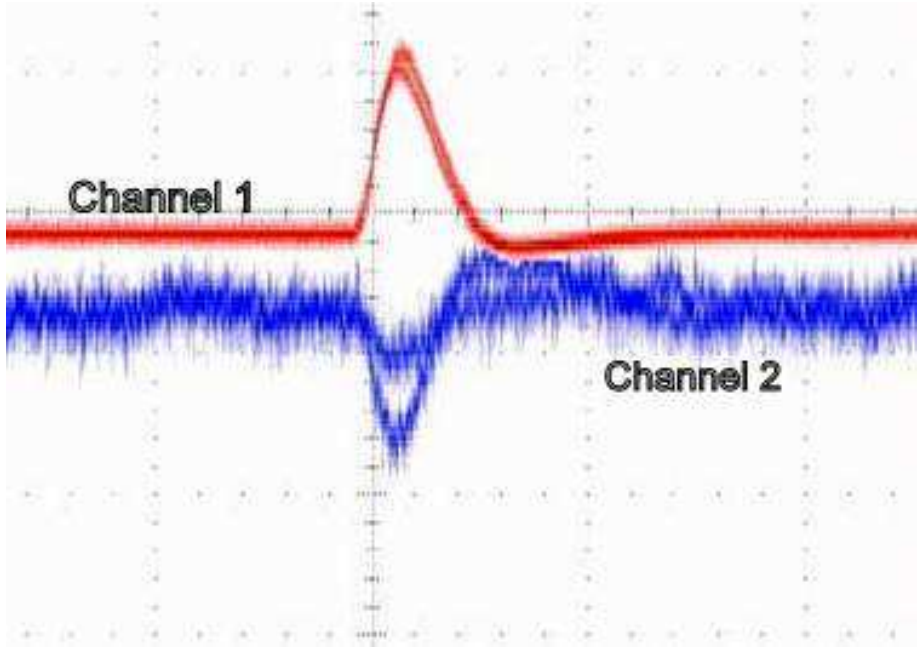}
  \caption{
\label{image}
Amplifier  (negative) outputs in Channel 2 corresponding to Image
pulses (negative) generated in the neighbouring segment (say,  2)
are   observed   simultaneously  if  positive  amplifier  outputs
corresponding to the real pulses in the segment number  1  (say,)
are used as the triggers in the oscilloscope.}

\end{figure}

Earlier  authors  have  discussed \cite{coo:1,coo:2}, that double
peaking is observed when addback spectra are  reconstructed  from
the  two-fold  events in a composite planar detector. However, in
their  work  \cite{coo:1}  ,  the  second  peak  originated  from
cross-talk  between two neighbouring segments. Cross talk induced
second peak usually appears at  higher  energy  compared  to  the
original one, due to its positive nature. However, in the present
case,  the second peak is at lower energies (Fig. \ref{twofold}),
indicating effect of image charge formation.

\subsubsection{Image charge formation}

For  segmented detectors, along with a net charge signal from the
collecting  electrode,  one  also  gets  transient   signals   of
neighboring  segments  where  image  charges  are  induced. These
transient signals are utilised to get  position  information  of
the interaction point of the gamma in the detector. The transient
pulses  corresponding  to image charge formation, are of opposite
polarity to the original pulses.  The  formation  of  the  pulses
corresponding  to image charges is evident when the pulses in two
segments of the LEPS detector are observed simultaneously if  one
is triggering the other, in the oscilloscope. Such characteristic
of  a  segmented  planar detector has also been discussed by R.J.
Cooper {\it et al.}. \cite{coo:1,coo:2}.  For  Compton  scattered
events which are detected in two neighbouring crystals, their sum
peak  may have peak height lower than the actual gamma energy due
to the presence of simultaneous negative pulses  originated  from
image  charges.  The variation of the heights of the image pulses
with change in the initial interaction position of the  gamma  in
the  detector  has  been  important  in developing imaging and or
tracking detectors \cite{coo:2}.

 \begin{figure}[htb]
\includegraphics[width=.8\linewidth,angle=0] {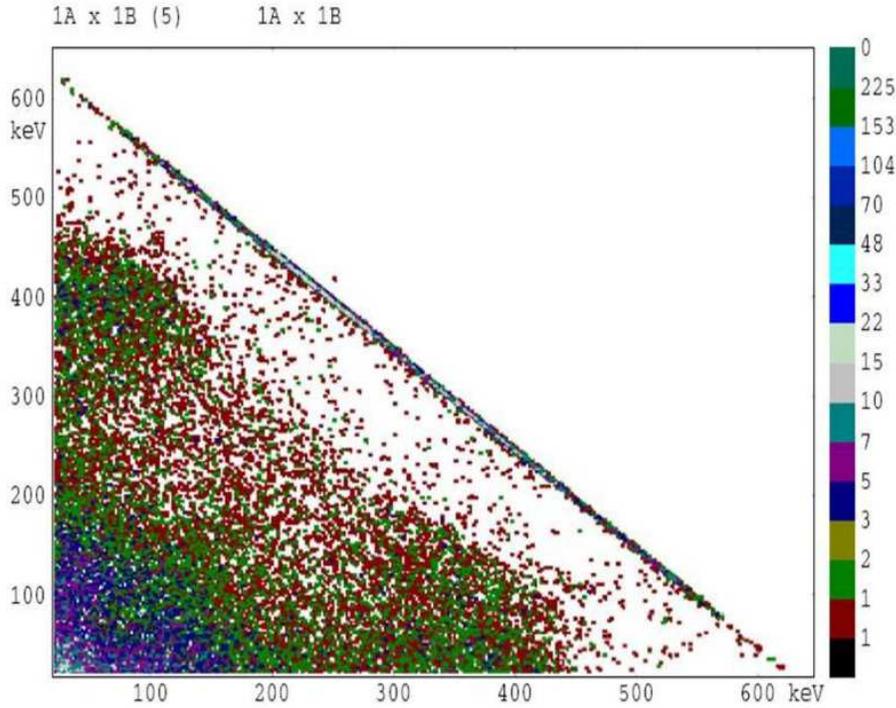}
\vspace{.8cm}
\caption{
\label{twod}
A  gamma- gamma two dimensional spectrum showing the simultaneous
events recorded by two crystals of the detector for 662 keV gamma
from a $^{137}Cs$ source.} \end{figure}

A  two dimensional matrix have been generated from the correlated
events with a $^{137}Cs$ source, and  is  plotted  in  Fig.6.  It
shows  an  intense diagonal line running from the top left to the
bottom right corresponding to true Compton  scattered  correlated
gamma  events detected in two adjacent segments which add up to a
total energy of 662 keV.

\begin{figure}[b]
\vspace{3. cm}
\includegraphics[width=.8\linewidth,angle=0] {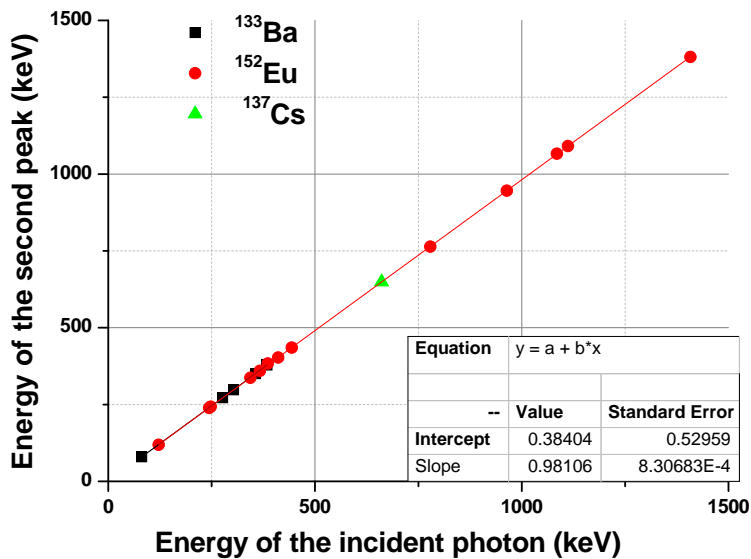}
\vspace{-3.8cm}
\caption{
\label{shift}
Plot  showing the energy of the second shifted (shadow) peak as a
function of energy of the original peak. } \end{figure}

\subsubsection{Correction}

The  magnitude  of the shift of the shadow peak originated due to
image charges was calculated by plotting the energy of the second
peak (in keV) as a function of gamma ray energy. This  result  is
displayed in Fig. \ref{shift}, showing the proportionality nature
of  the  shift. The gradient of the fitted line indicates that the
magnitude of this shift is only $\simeq  2\%$  of  the  deposited
energy.

\begin{figure}[h]
\vspace{5. cm}
\includegraphics[width=1.\linewidth,height=.7\linewidth,angle=0]
{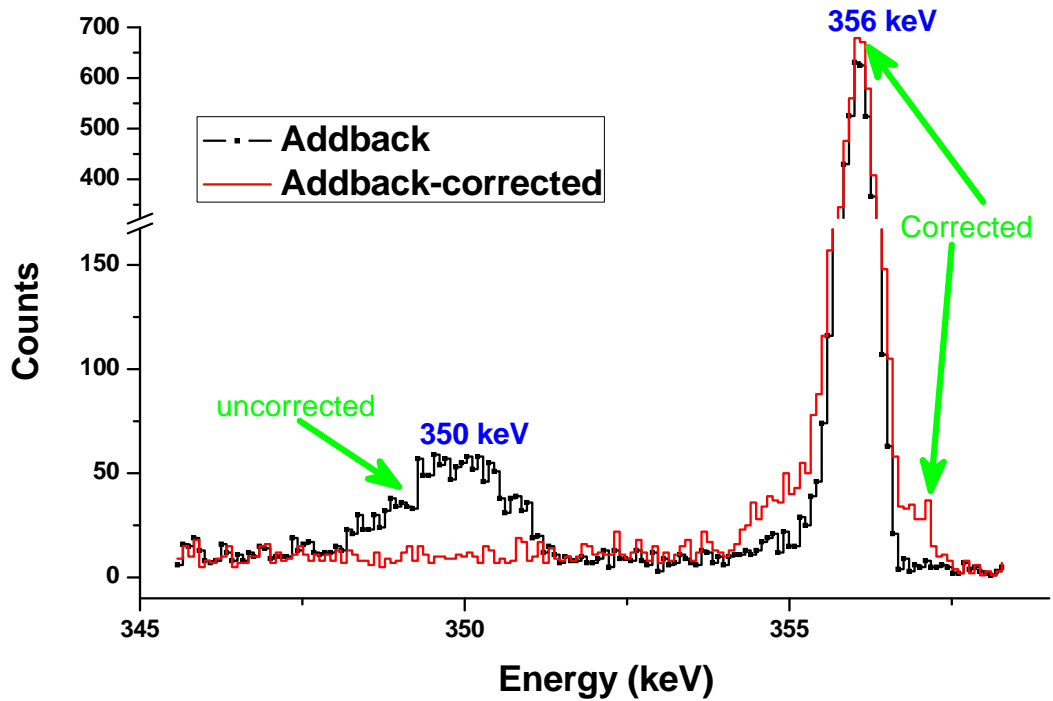}
\vspace{-5cm}
\caption{
\label{correc}
Spectra  showing  the  356  keV  photopeak from $^{133}Ba$ in the
addback spectra before (contains a second shadow  peak  at  lower
energies) and after shift correction.} \end{figure}

This  energy shift in the peak generated from the two fold events
can be compensated by applying an event by  event  correction  in
the  calculated addback energy. The effect of this correction can
be seen in Fig. \ref{correc}. It  is  seen  that  the  correction
removes  the  events from the shadow peak to improve the count of
the full energy peak. The full energy peak for 356 keV gamma from
a $^{133}Ba$ source has been compared  with  the  same  photopeak
after  reconstruction  including  the corrected two - fold events
(Fig. \ref{correc}). However, at higher energies,  gradually  the
base  of the peak widens and the FWTM (full width at one-tenth of
maximum) deteriorates after correction. This is due to the  shift
variation  according to the relative location of the two segments
of the detectors fired simultaneously. Thus, the corrections need
to be different depending  on  this  relative  position.  In  the
present  case,  shown  in Fig. \ref{correc}, different pairs were
not distinguished and the same correction factor was used for any
two -fold event depending on the incident energy.  The  variation
of  the  heights  of  the image pulses with change in the initial
interaction position of the gamma  needs  to  be  considered  for
better reconstruction of the spectrum.

 \subsubsection{Addback Factor}

\begin{figure}[ht]
\vspace{3. cm}
\includegraphics[width=.8\linewidth,height=.7\linewidth,angle=0]
{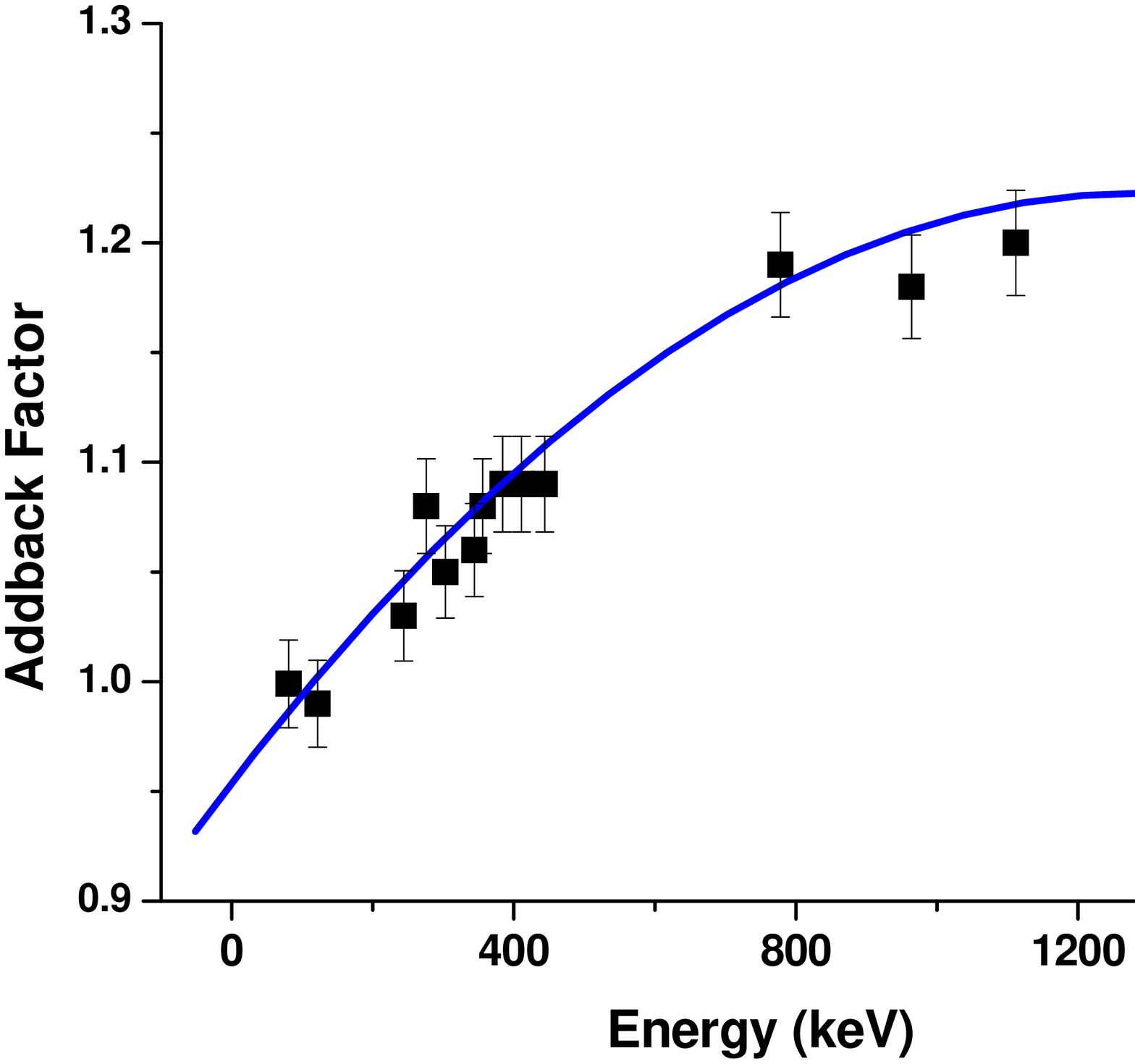}
\vspace{-3.8cm}
\caption{
\label{addback}
Variation of the addback factor as a function of energy, after
shift correction.}
\end{figure}

This  correction factor improved the performance of the composite
LEPS detector which has been estimated by the addback  factor  as
usually defined in composite detectors \cite{mss1}. An estimation
of  the variation of the addback factor with increase in incident
gamma energy is shown in Fig. \ref{addback}.

 \section{Conclusion}
In  the  present study, the composite LEPS has been characterised
using radioactive sources up  to  1.4  MeV.  Formation  of  image
charges  in neighbouring segments deteriorates the performance of
the detector. Event-by-event correction has been implemented. The
corrected results indicate that these detectors can also be  used
in  addback  mode with an increase in efficiency by $\simeq 20\%$
at  1.408  MeV.  However,  to  improve  the  peak  shape  of  the
reconstructed  peak, one may use different correction factors for
different relative locations of the two segments of the detectors
fired simultaneously.

\acknowledgments

The  authors  sincerely thank Ms. Jonaki Panja for technical help
during the experiment.

\end{document}